\documentclass[twocolumn]{IEEEtran}
\usepackage{amsmath}
\usepackage{amssymb}
\usepackage{amsfonts}
\usepackage{color}
\usepackage{graphicx}
\usepackage{subcaption}
\usepackage{mathrsfs}
\usepackage{times}
\usepackage{empheq}
\usepackage{epstopdf}
\usepackage{cite}
\usepackage{relsize}
\usepackage{relsize}
\usepackage{arydshln}
\usepackage{float}
\usepackage[normalem]{ulem}
\usepackage{mathtools,amssymb,lipsum}
\usepackage{cuted}
\usepackage{lettrine}

\title{Interference Analysis of QAM based Filter Bank Multicarrier System with Index Modulation}
\author{Adnan Zafar, Aijun Cao, Mahmoud Abdullahi, Lei Zhang, Pei Xiao and Muhammad Ali Imran
	\thanks{A.Zafar, M.Abdullahi and P.Xiao are with Institute for Communication Systems (ICS), University of Surrey, Guildford, UK. A.Zafar is also with Institute of Space Technology, Islamabad, Pakistan. Email: \{a.zafar, m.abdullahi, p.xiao\}@surrey.ac.uk}
	\thanks{A.Cao is with ZTE R\&D Center, 164 51 Stockholm, Sweden. Email: cao.aijun@zte.com.cn}
	\thanks{L.Zhang and M.A.Imran are with School of Engineering, University of Glasgow, Glasgow, UK. Email: \{lei.zhang, muhammad.imran\}@glasgow.ac.uk}
}

\begin{document}
\maketitle
\begin{abstract}
Index modulation (IM) has recently emerged as a promising concept for spectrum and energy-efficient next generation wireless communications systems since it strikes a good balance among error performance, complexity, and spectral efficiency. IM technique when applied to a multicarrier waveforms yields the ability to convey the information not only by $\mathcal M$-ary signal constellations as in conventional multicarrier systems, but also by the indexes of the subcarriers, which are activated according to the incoming bit stream. Although IM is well studied for OFDM based systems, FBMC with index modulation has not been thoroughly investigated. In this paper, we shed light on the potential and implementation of IM technique for QAM based FBMC system. We start with a mathematical model of the IM based QAM-FBMC system (FBMC/QAM-IM) along with the derivation of interference terms at the receiver due to channel distortions and noise. The interference terms including the ones introduced by the multipath channel are analyzed in terms of MSE and output SINR. It is shown with analytical and simulation results that the interference power in FBMC/QAM-IM is smaller comparing to that of the conventional FBMC/QAM system as some of the subcarriers are inactive. The performance of FBMC/QAM with IM is investigated by comparing the SIR and output SINR with that of the conventional FBMC/QAM system along with the BER performance which shows that the FBMC/QAM-IM is a promising transmission technique for future wireless networks.
\end{abstract}
\section{Introduction}
\lettrine{I}{ndex} modulation technique, which utilizes an innovative way of conveying information compared to conventional communication systems, has emerged as a promising candidate scheme for 5G wireless networks \cite{Wen2017}. In this scheme, information is not only conveyed by the $\mathcal M$-ary signal constellation, but also by the indexes of the building blocks of the corresponding communication system. The two main promising applications of IM is in multiple antenna systems like MIMO and multicarrier schemes like OFDM. In MIMO, IM can be applied on the transmit antennas of the MIMO system to convey additional information. This application of IM to MIMO systems is termed as SM \cite{4382913}. On the other hand application of IM with a multicarrier scheme like OFDM considers subcarrier indexes as a source of conveying additional information \cite{7469311}. IM based OFDM (OFDM-IM) system can also be considered as a frequency domain extension of the SM concept. Recent studies have shown that OFDM-IM can offer appealing advantages over classical OFDM and is also being proposed as a strong waveform candidate for future wireless networks \cite{8004416}. The performance improvement in OFDM-IM system comes from the fact that the system utilizes the indexes of the active subcarriers to convey additional information bits. At the transmitter side of the OFDM-IM system, the subcarriers in each symbol are divided in to sub-groups comprising of active and inactive subcarriers. The indexes of these active subcarriers are used as a source of additional information which may leads to improvement in the transmission efficiency of the system. At the receiver side, the active subcarrier indexes can either be detected by a ML detector or a low-complexity LLR detector \cite{6503868}. Furthermore, in IM based multicarrier systems, the power of inactive subcarriers can either be reallocated to active subcarriers in each sub-group to improve the BER performance of the system and can also be saved to improve the energy efficiency of the system \cite{5449882}. This has enabled IM to emerge as a promising spectral and energy efficient modulation schemes for future wireless communications. A generalization of OFDM-IM have also been proposed in recent years to achieve higher spectral efficiencies \cite{7112187} and improved system performance \cite{6841601}. A direct combination of OFDM-IM with MIMO transmission techniques, called as MIMO-OFDM-IM, is proposed in \cite{7234862,7448967}. It is shown that the proposed MIMO-OFDM-IM system can achieve a linear increase in spectral efficiency (SE) of the system. Despite several advantages, conventional OFDM based systems have various short comings including poor OoBR and strict synchronization requirements.\\
In recent years, FBMC/OQAM has emerged as a promising candidate waveform for future wireless networks due to its profound advantages over conventional OFDM \cite{7390479,8100961,7549072}. However, conventional FBMC/OQAM suffers from pure imaginary intrinsic interference caused by neighboring symbols. Some initial work on the application of IM with FBMC/OQAM (FBMC/OQAM-IM) has been investigated and it is shown that FBMC/OQAM-IM has improved performance compared to conventional FBMC/OQAM system. However, when the IM scheme is applied to FBMC/OQAM system, the intrinsic interferences are partly eliminated and the remaining interferences still affect the BER performance \cite{7873550,8287935,8171193}. Although IM is well studied for OFDM based systems and some literature on FBMC with OQAM modulation is available, FBMC/QAM with index modulation (FBMC/QAM-IM) has not been thoroughly investigated.
\begin{figure*}[t]
	\centering
	\includegraphics[scale=0.6]{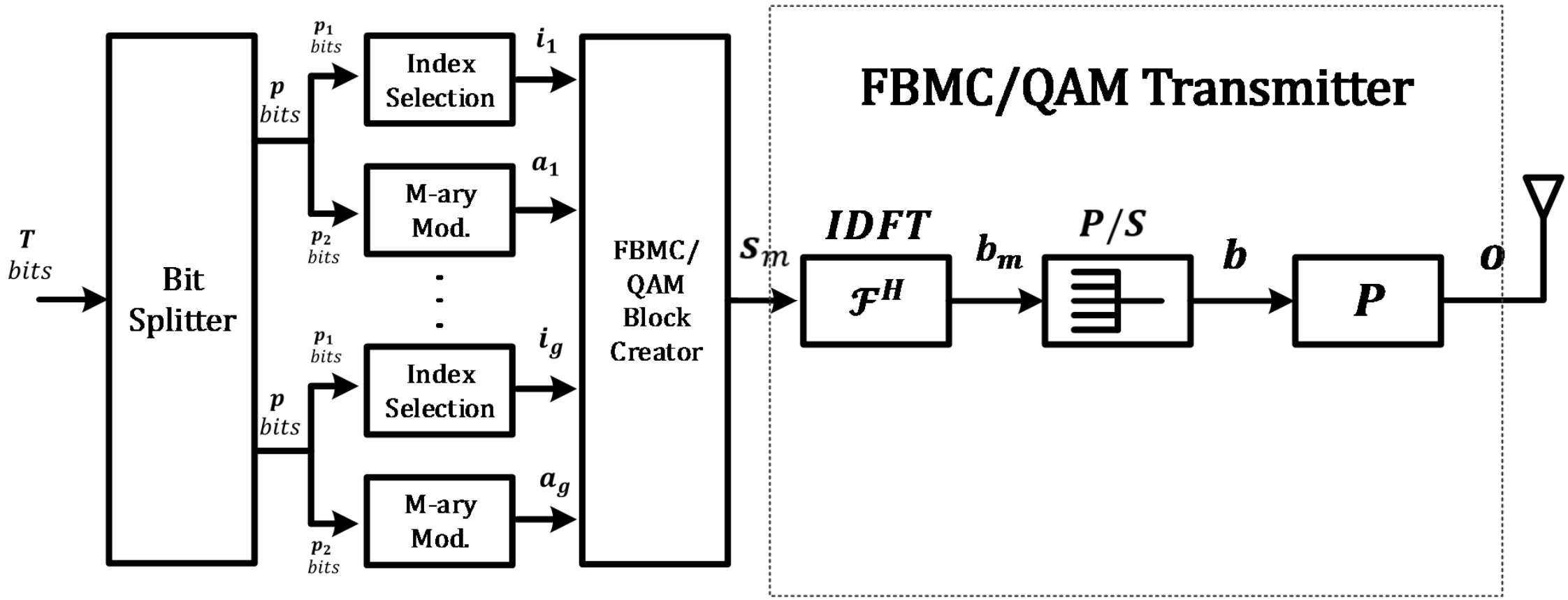}
	\caption{Block diagram of FBMC/QAM-IM Transmitter}
	\label{fig-5.1}
\end{figure*}
\\\\The main work in this contribution is summarized as follows
\begin{itemize}
	\item A mathematical matrix model of the index modulation based FBMC/QAM system is presented along with the derivation of interference terms at the receiver due to channel distortions and the intrinsic behavior of the transceiver model.
	\item The interference terms including the ones introduced by the multipath channel are analyzed in terms of MSE with and without IM. It is analytically shown that the interference power in FBMC/QAM-IM is smaller comparing to that of the conventional FBMC/QAM system as some subcarriers are inactive.
	\item The performance of FBMC/QAM-IM is evaluated by comparing the SIR and output SINR with that of the conventional FBMC/QAM system.
	\item An improved LLR detector is then proposed based on the proposed interference model. It is shown with simulations that the proposed detector provide improved BER performance in FBMC/QAM-IM system compared to the conventional FBMC/QAM system.
\end{itemize}

The rest of this paper is organized as follows: \\
A mathematical model of the IM based QAM-FBMC system is presented in Section II. The interference terms including the ones introduced by the multipath channel and the intrinsic behavior of the FBMC/QAM transceiver model are analyzed in terms of MSE with and without IM and are presented in Section III. We proposed an improved LLR detector based on the interference analysis in Section IV. The simulation results including MSE, SIR, output SINR and BER are provided in Section V. Lastly, conclusion is provided in Section VI.
\paragraph*{Notations} Vectors and matrices are denoted by lowercase and uppercase  bold letters. $\lfloor$x$\rfloor$ represent the greatest integer smaller than x. $\{\cdot\}^H, \{\cdot\}^T, \{\cdot\}^*$ stand for the Hermitian conjugate, transpose and conjugate operation, respectively. $\mathcal {E}\{\mathbf A\}$ denotes the expectation operation of $\mathbf A$. $\mathcal {F}$ and $\mathcal F^H$ represents the power normalized N point discrete Fourier transform (DFT) and inverse DFT matrices. $\mathbf I_{m\times m}$ refers to $m$ dimension identity matrix and for some cases the subscript will be dropped for simplification. $A\otimes B$ represents kronecker product of A and B. $\|\mathbf A\|_n^2$ means taking the $n^{th}$ diagonal element of matrix $\|\mathbf A\|^2=\mathbf A \mathbf A^H$. 
We use $*$ as a linear convolution operator. $\textrm{Tr}\{\mathbf A\}$ denotes the trace of matrix $\mathbf A$.
\section{FBMC/QAM System with Index Modulation}\label{sec5.2}
In this section we define the index modulation based FBMC/QAM system in matrix form which will be subsequently used to analyze the interference reduction due to the introduction of IM in conventional FBMC/QAM system. 
\subsection{System Model}
The system model of FBMC/QAM-IM is developed as an extension to the FBMC/QAM model presented in \cite{2017arXiv171108850Z}. We will focus on the index modulation block in detail while the FBMC/QAM model will be briefly discussed. 
\subsubsection{Index Modulation}
The block diagram of the FBMC/QAM-IM system is illustrated in Fig. {\ref{fig-5.1}}. Since, index modulation utilizes indexes of certain subcarriers according to the incoming bit stream to convey information, we define the subcarriers bearing $\mathcal M$-ary signal constellation symbols as active subcarriers and the rest as inactive. The proposed model of FBMC/QAM-IM follows a block based processing approach where each block contains $M$ symbols and each symbol has $N$ subcarriers in the frequency domain. 
Let us consider that transmitted information of each block has $T$ bits. These information bits are then divided into $g$ groups each containing $p$ bits which are then mapped to an FBMC/QAM subblock of length $n$, where $n=N/g$. Each group of $p$ bits can be split into two parts. The first $p_1$ bits are used for index selection in which $k$ out of $n$ available subcarrier are activated according to a predefined look up table \cite{6503868}. The remaining $p_2$ bits are mapped on to the $\mathcal M$-ary signal constellation to define the data symbols that will modulate the active subcarriers. The number of bits $p_1$ carried by the indexes of the active subcarriers are defined as 
\begin{eqnarray}\label{5.1}
p_1 = \lfloor \log_2 C^n_k\rfloor,
\end{eqnarray}
where $C^n_k$ denotes the binomial coefficient. In other words, selected indexes can have $c=2^{p_1}$ possible realizations. The number of bits carried by the $\mathcal M$-ary constellation symbols are defined as 
\begin{eqnarray}\label{5.2}
p_2 = k\log_2 \mathcal M,
\end{eqnarray}
The total number of bits that are transmitted by one FBMC/QAM-IM block can be defined as
\begin{eqnarray}\label{5.3}
T &=& (p_1+p_2)g \nonumber \\
&=&(\lfloor \log_2 C^n_k\rfloor+k\log_2 \mathcal M)g
\end{eqnarray}
In other words, in FBMC/QAM-IM system, the information is conveyed by both the indexes of the active subcarriers and also the $\mathcal M$-ary constellation symbols modulating these active subcarriers. Also we can infer that, as we are not using all of the available subcarriers for the data transmission, the loss in transmission efficiency is compensated by transmitting additional bits in the spatial domain of the FBMC/QAM block. From Fig. {\ref{fig-5.1}}, it can be seen that for each subblock $\beta$, the first $p_1$ bits are used for index selection i.e., $k$ out of $n$ available indexes are selected as 
\begin{eqnarray}\label{5.4}
{\bf{\textit{i}}}_\beta = [i_{\beta,1},...,i_{\beta,k}]
\end{eqnarray}
where $i_{\beta,\gamma} \in [1,...,n]$, $\beta=1,...,g$ and $\gamma=1,...,k$. To modulate these active subcarriers, the remaining $p_2$ bits are mapped on the $\mathcal M$-ary signal constellation to define the transmitted data symbols as
\begin{eqnarray}\label{5.5}
{\bf{\textit{a}}}_\beta = [a_{\beta,1},...,a_{\beta,k}]
\end{eqnarray}
where $a_{\beta}$ is the symbol vector transmitted on the active sub-carriers in each sub-block. Such that $a_{\beta,\gamma}\in \mathcal{S}$ for $\beta=1,...,g$ and $\gamma=1,...,k$, where $\mathcal{S}$ is the set of all possible complex symbols from the $\mathcal M$-ary constellation. Using ${\bf{\textit{i}}}_\beta$ and ${\bf{\textit{a}}}_\beta$ from (\ref{5.4}) and (\ref{5.5}), the FBMC/QAM block creator generates all of the subblocks $\mathbf A_{m,\beta}\in\mathbb{C}^{n\times 1}$ for $\beta=1,...,g$ and form FBMC/QAM-IM symbol as ${\mathbf{s}_m}=[{s_{m,0}},{s_{m,1}},...,{s_{m,N-1}}]^T\!\!=\!\![{\mathbf A_{m,1}},{\mathbf A_{m,2}},{...},{\mathbf A_{m,g}}]^T\in \{0,a_{\beta,\gamma}\}$. The generated FBMC/QAM-IM block can be expressed as ${\mathbf S}\!\!=\!\![{\mathbf{s}_0},{\mathbf{s}_1},{\mathbf{...}},{\mathbf{s}_{M-1}}]\in \mathbb{C}^{N\times M}$. Furthermore, to have unit average power, we reallocate the power from inactive subcarriers to the active subcarriers in a subgroup. The average power of the modulated symbol $s_{m,n}$ can be represented as $\mathcal{E}\{\|s_{m,n}\|^2\}= \delta^2$. It should be noted that when the number of active subcarriers $k$ is equal to the number of carriers in a subgroup i.e., $k=n$, the FBMC/QAM-IM system will become a conventional FBMC/QAM system. Hence, the FBMC/QAM system can be viewed as a special case of the FBMC/QAM-IM system.\\
It should be noted that by properly choosing $n$ and $k$, the spectral efficiency of FBMC/QAM-IM system can be improved. The spectral efficiency $\eta_{FBMC/QAM-IM}$ can be calculated as follows
\begin{eqnarray}\label{5.6}
\eta_{FBMC/QAM-IM}\!\! =\!\! \frac{M \frac{N}{n}(\!\lfloor\! \log_2\! C^n_k\!\rfloor\!\!+\!\!k\log_2\! \mathcal M\!)}{NM} \textit{bits/sec/Hz}
\end{eqnarray}
As we already mentioned that FBMC/QAM-IM system reduces to FBMC/QAM system in case of $k=n$, the spectral efficiency of FBMC/QAM can now be derived using (\ref{5.6}) as
\begin{eqnarray}\label{5.7}
\eta_{FBMC/QAM} = \log_2 \mathcal M\quad \quad \textit{bits/sec/Hz}
\end{eqnarray}
The spectral efficiency gain of FBMC/QAM-IM over conventional FBMC/QAM system can be expressed as
\begin{eqnarray}\label{5.8}
\eta_{gain} = \frac{\eta_{FBMC/QAM-IM}}{\eta_{FBMC/QAM}}\!\!\!\!\!&=&\!\!\!\!\!\frac{M \frac{N}{n}(\lfloor \log_2 C^n_k\rfloor+k\log_2 \mathcal M)}{NM\log_2 \mathcal M}\nonumber\\
\!\!\!\!\!&=&\!\!\!\!\! \frac{\lfloor \log_2 C^n_k\rfloor}{n\log_2 \mathcal M}+\frac{k}{n}
\end{eqnarray}
We can see from (\ref{5.8}), that by properly selecting $n$, $k$ and $\mathcal M$-ary modulation, we can achieve spectral efficiency gain $\eta_{gain} > 1$. For example, when $n=4$, $k=3$ and $\mathcal M = 2$ i.e., BPSK modulation, we can calculate $\eta_{gain}=1.25$, which indicates that the spectral efficiency of FBMC/QAM-IM has exceeded that of conventional FBMC/QAM.
\subsubsection{Transmit Processing}
It can be seen from Fig. {\ref{fig-5.1}} that after index modulation, the signal $\mathbf{{s}}_m$ is passed though the conventional FBMC/QAM transmitter i.e., $N$ point IDFT processor, parallel to serial conversion and the transmit filter matrix $\mathbf P$. The structure of $\mathbf P$ is already defined in \cite{2017arXiv171108850Z}. The output of the transmit filter matrix can be expressed as 
\begin{eqnarray}\label{5.9}
{\mathbf{o}}={\mathbf{P}{\mathbf{b}}}\in \mathbb{C}^{(K+M-1)\times1},
\end{eqnarray}
where $\mathbf {b}$ is the signal vector processed symbol by symbol through the IDFT block i.e., $\mathbf{\mathcal{F}}^H$. The signal vector can be represented as ${\mathbf{{b}}}=[{\mathbf{{b}}_0};{\mathbf{{b}}_1};{\mathbf{\cdots}};{\mathbf{{b}}_{M-1}}]=[{\mathbf{\mathcal{F}}^H}{\mathbf{{s}}_0};{\mathbf{\mathcal{F}}^H}{\mathbf{{s}}_1};{\mathbf{\cdots}};{\mathbf{\mathcal{F}}^H}{\mathbf{{s}}_{M-1}}]\in \mathbb{C}^{MN\times1}$. It should be noted that the prototype filter matrix  $\mathbf{P}$ is designed in a manner that when it is multiplied by vector $\mathbf{b}$; the multiplication of matrices is equivalent to the required linear convolution process. As a result the output of the transmit filter will have $(K-1)N$ more samples than the signal vector $\mathbf b$ as can be seen from (\ref{5.9}).
\subsubsection{Passing through the Channel}
We assume the system operates over a slowly-varying fading channel i.e., quasi-static fading channel. In such a scenario, we can assume that the duration of each of the transmitted data block is smaller than the coherence time of the channel, therefore the random fading coefficients stay constant over the duration of each block \cite{tse2005fundamentals}. In this case, we define the multipath channel as a $L$-tap channel impulse response (CIR) with the $l^{th}$-tap power being ${\rho^2_l}$. It is also assumed that the average power remains constant during the transmission of the whole block. Let us define the CIR as
\begin{eqnarray}\label{5.10}
\mathbf h &=& [h_{0},h_{1},\cdots,h_{L-1}]^T\nonumber \\
&=& [\rho_0 z_{0},\rho_1 z_{1},\cdots,\rho_{L-1} z_{L-1}]^T,
\end{eqnarray}
where $h_{l}$ denotes the channel coefficient of the $l^{th}$ tap in the time domain and the complex random variable $z_l$ with complex Gaussian distribution as $\mathbb{C}\mathcal{N}(0,1)$ represents the multipath fading factor of the $l^{th}$ tap of the quasi-static rayleigh fading channel. The signal vector $\mathbf{o}$ is then passed through the multipath channel $\mathbf h$ as discussed in \cite{2017arXiv171108850Z}. The received signal after passing through the channel can be represented as
\begin{eqnarray}\label{5.11}
\mathbf  {r} = \sum_{l=0}^{L-1} \rho_l \mathbf Z_{l}\mathbf{P}\mathbf{b}_e^{\downarrow l} + \mathbf o_{fd}+ \mathbf o_{IBI}+\mathbf n,
\end{eqnarray}
The first term in (\ref{5.11}) represents the linear convolution process between the channel and the transmitted signal where $\mathbf Z_{l}$ implies that each FBMC symbol in a block experiences the same channel i.e., $\mathbf{Z}_{l}= z_{l}\times \mathbf I_{(K+M-1)N \times (K+M-1)N}$ and $\mathbf{b}^{\downarrow l}_e \!=\! \mathbf {X}_{l} \mathbf{b}$. Note that the matrix $\mathbf X_{l}\!=\! I_M \otimes \mathbf {X}_{sub,l}$ is a block diagonal exchange matrix where $\mathbf {X}_{sub,l} \!=\![\mathbf {0}_{l\times (N-l)}\!\!\quad\!\!,\!\!\quad\!\! \mathbf {I}_{l} \!\!\quad\!\!;\!\!\quad\!\!  \mathbf {I}_{N-l}\!\!\quad\!\!,\!\!\quad\!\! \mathbf {0}_{(N-l)\times l}]$ and is used to exchange the locations of elements of $\mathbf{b}$. Furthermore, $\mathbf o_{fd}=\sum_{l=0}^{L-1} \rho_l \mathbf Z_{l}\Delta\mathbf{P}^{\downarrow l}\mathbf{b}_e^{\downarrow l}$ is the interference caused by the filter distortion due to channel multipath effect, $\mathbf o_{IBI}=\sum_{l=0}^{L-1}\rho_l \mathbf Z_l\mathbf r_{B,l}$ is the inter-block interference (IBI) caused by multipath channel with $\mathbf r_{B,l} = [\mathbf r_{p,l}; \mathbf 0_{[(M+K-l)N-l]\times 1}]$ and $\mathbf r_{p,l} \in \mathbb{C}^{l\times 1}$ is the interfering signal from the previous FBMC/QAM block and $\mathbf{n}$ is a Gaussian noise vector with each element having zero mean and variance $\sigma^2$.
\begin{figure*}[t]
	\centering
	\includegraphics[scale=0.6]{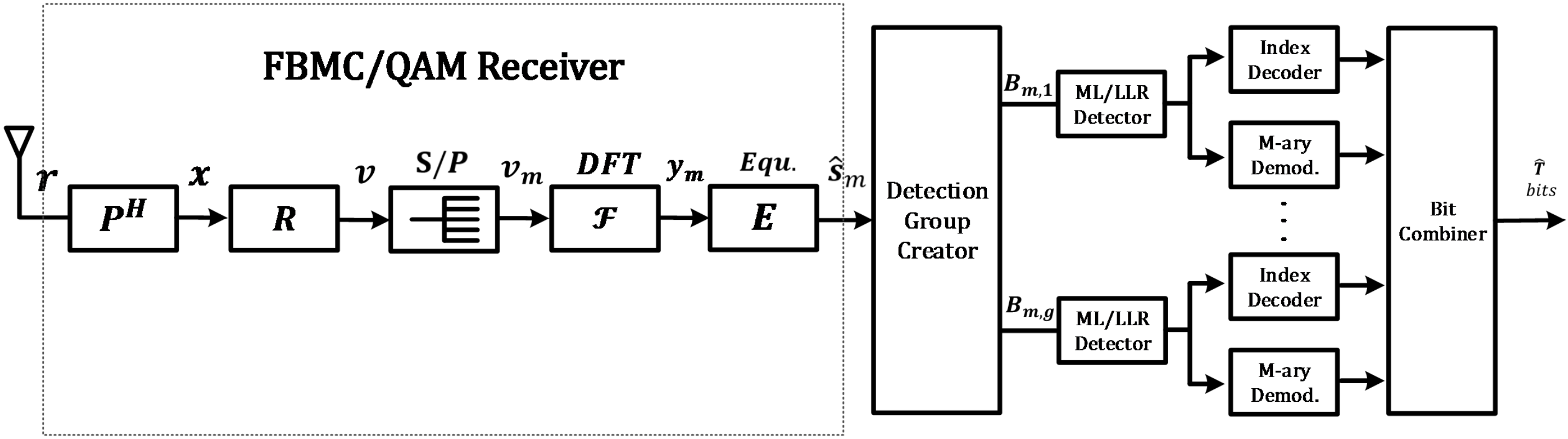}
	\caption{Block diagram of FBMC/QAM-IM Receiver}
	\label{fig-5.2}
\end{figure*}
\subsubsection{Receive Processing}
The receiver process of FBMC/QAM-IM is shown in Fig. \ref{fig-5.2}. At the receiver, the received signal $\mathbf r$ is first passed through the receive filter $\mathbf {P}^H$, which serve as a matched filter in FBMC/QAM receiver. The signal after matched filtering is then passed though the inverse filter $\mathbf R$ to cancel the intrinsic interferences in the received signal block \footnote{The concept of using inverse filter in FBMC/QAM system to cancel the self-interference (so called intrinsic interference) has been proposed in \cite{2017arXiv171108850Z}}. The signal after the inverse filtering is then processed symbol by symbol through the DFT processor i.e., $\mathcal F$. The resulting frequency domain signal $\mathbf y_m$ is then equalized using ZF or MMSE equalizer $\mathbf E$ to counter the effects of multipath channel. The complete receive processing is already derived in detail in \cite{2017arXiv171108850Z}. Following the same receiver processing steps, the equalized symbol $\mathbf {\hat s}_m$ can now be represented as follows
\begin{eqnarray}\label{5.12}
\mathbf {\hat s}_m&=&\underbrace{\mathbf s_m}_{\textrm{Desired Signal}}\!\!+\!\!\underbrace{(\mathbf{I}-\boldsymbol{\beta})\mathbf s_m}_{\textrm{MMSE Estimation Bias}}+\!\!\! \underbrace{\mathbf E\mathcal{F}\mathbf R_m\mathbf P_m^H \mathbf o_{fd}}_{\textrm{Filter Distortion by Multipath}}\!\!\! \nonumber \\&&+ \underbrace{\mathbf E\mathcal{F}\mathbf R_m\mathbf P_m^H\mathbf o_{IBI}}_{\textrm{IBI by Multipath}}+ \underbrace{\mathbf E\mathcal{F}\mathbf R_m\mathbf P_m^H\mathbf n}_{\textrm{Noise}}
\end{eqnarray}
\indent  
where $\boldsymbol{\beta} = \mathbf E\mathbf C$ is a diagonal matrix with its $n^{th}$ diagonal element being defined as
\begin{eqnarray}\label{eq:26}
{\beta}_{n} = {E}_{n} {C}_{n}= \frac{|E_{n}|^2}{|E_{n}|^2 + \nu\sigma^2/\delta^2},
\end{eqnarray}
and $\mathbf C$ is the frequency domain channel coefficients in diagonal matrix form and is given as $\mathbf C=\textrm{diag}[\mathbf C_{0},\mathbf C_{1},\cdots,\mathbf C_{N-1}] \in \mathbb{C}^{N \times N}$. The $n^{th}$ block diagonal element in the frequency response of the channel can be represented as $\mathbf C_{n}=\sum_{l=0}^{L-1} h_{l}e^{-j\frac{2\pi}{N}nl},0\le n\le N$.
As we can see from (\ref{5.12}) that the transmitted signal $\mathbf s_m$ is free from ICI and ISI terms due to the use of inverse filter matrix $\mathbf R$. However, the received symbols are affected by interferences caused by MMSE estimation bias, IBI and filter distortion due to multipath channel. Since, we know that some of the sub-carriers in FBMC/QAM-IM system are in-active and therefore they will not contribute to the overall interference in the system. In what follows, we will evaluate the interference and noise power in FBMC/QAM and FBMC/QAM-IM systems to estimate the performance improvement in term of MSE and output SINR due to the introduction of index modulation.
\section{Interference and Noise Power Estimation}\label{sec5.3}
It can be seen from (\ref{5.12}), that the transmitted symbol vector $\mathbf s_m$ is accompanied with interference terms caused by the multipath channel and noise. In this section we estimate the interference and noise power in the FBMC/QAM-IM system. First, let us introduce a diagonal matrix $\mathbf D_m\in \mathbb R^{N\times N}$ such that the $n^{th}$ diagonal element of $\mathbf D_m$ is defined as 
\begin{eqnarray}\label{5.13}
D_{m,n} = \left\{
\begin{array}{lcl}
1 \quad \quad \textrm{if} \quad  s_{m,n}\in a_{\beta,\gamma}\\
0 \quad \quad \textrm{if} \quad  s_{m,n}=0\\
\end{array}
\right.
\end{eqnarray}
The diagonal matrix $\mathbf D_m$ represent the active and inactive indexes of the transmitted symbol vector $\mathbf s_m$. 
It should be noted that introducing $\mathbf D_m$ will have no effect on the system model since $\mathbf D_m \mathbf s_m = \mathbf s_m$. The purpose of defining $\mathbf D_m$ is to evaluate the impact of inactive subcarriers on the system performance. Since some of the subcarriers in FBMC/QAM-IM are inactive, therefore they will not contribute to the overall interference in the system. Furthermore, it should be noted that $\mathbf D_m$ will be an identity matrix in case of conventional FBMC/QAM system since all of the subcarriers are active i.e., $n=k$ and contribute to the overall interference in the system. It can be seen from (\ref{5.12}) that the estimated symbol is accompanied with MMSE estimation bias, interference terms like filter distortion and IBI due to multipath channel and noise i.e.,
\begin{eqnarray}\label{5.14}
\mathbf {\hat s}_m&=&\underbrace{\mathbf s_m}_{\textrm{Desired Signal}}+\underbrace{\boldsymbol{\psi}_{resd,m}}_{\textrm{MMSE Estimation Bias}}+\!\!\! \underbrace{\boldsymbol{\psi}_{fd,m}}_{\textrm{Filter Distortion by Multipath}}\!\!\!\nonumber \\&&+ \underbrace{\boldsymbol{\psi}_{IBI,m}}_{\textrm{IBI by Multipath}}+ \underbrace{\boldsymbol{\psi}_{noise,m}}_{\textrm{Noise}},
\end{eqnarray}
\indent The MSE of the $n$-th modulation symbol estimation in the $m$-th FBMC/QAM-IM symbol can be derived as
\begin{eqnarray}\label{5.15}
\gamma_{tot,m,n}\!\!\!\! &=&\!\!\!\! \mathcal {E}||{\hat s}_{m,n}- {s}_{m,n}||^2 \nonumber \\
\!\!\!\!&=&\!\!\!\! \mathcal {E}\big[ \|\boldsymbol{\psi}_{resd,m}\|_n^2+ \|\boldsymbol{\psi}_{fd,m}\|_n^2+\|\boldsymbol{\psi}_{IBI,m}\|_n^2\nonumber \\
&&+\|\boldsymbol{\psi}_{noise,m}\|_n^2\big]
\end{eqnarray}
\subsection{Variance of signal estimation bias}
The desired signal estimation bias is caused by the MMSE receiver since it minimizes the MSE between the transmitted and received signal. This leads to residual interference in the estimated signal. From (\ref{5.15}) and (\ref{5.12}), we can write the variance of the signal estimation bias as
\begin{eqnarray}\label{5.16}
\boldsymbol\gamma_{resd,m,n}\!\!\!&=&\!\!\!\mathcal {E}\|\boldsymbol{\psi}_{resd,m}\|_n^2 = \mathcal {E}\{\|(\mathbf I -\boldsymbol\beta)\mathbf s_m\|_n^2\}\nonumber \\
&=&\mathcal {E}\{\|(\mathbf I -\boldsymbol\beta)\mathbf D_m\mathbf s_m\|_n^2\}\nonumber \\
&=&\!\!\!\delta^2\|(\mathbf I -\boldsymbol\beta)\mathbf D_m\|_n^2\nonumber \\
&=&\delta^2D_{m,n}(I -\beta_n)^2
\end{eqnarray}
where $\mathcal{E}\{\mathbf s_m\mathbf s_m^H\}\!\!=\!\! \delta^2\mathbf I$ and according to (\ref{eq:26}), ${\beta_n} = \frac{|C_n|^2}{|C_n|^2 + \nu\sigma^2/\delta^2}$. Substituting ${\beta_n}$ into (\ref{5.16}) yields
\begin{eqnarray}\label{5.17}
\gamma_{resd,m,n} \!\!\!\!\!&=&\!\!\!\!\! \delta^2D_{m,n}(I -\beta_n)^2=\delta^2D_{m,n}(I -2\beta_n+\beta_n^2], \nonumber \\
\!\!\!\!\!&=&\!\!\!\!\!\delta^2D_{m,n}\Big[\!I\!\! -\!\!\frac{2|C_n|^2}{|C_n|^2 \!+\! \nu\sigma^2/\delta^2}\!\!+\!\!\frac{|C_n|^4}{(|C_n|^2 \!+\! \nu\sigma^2/\delta^2)^2}\Big],\nonumber \\
\!\!\!\!\!&=&\!\!\!\!\!\delta^2D_{m,n}\bigg[\frac{\nu^2\sigma^4}{(\delta^2|C_n|^2 + \nu\sigma^2)^2}\bigg].
\end{eqnarray}
Apparently, when the ZF receiver is adopted, $\gamma_{resd,m,n} = 0 $ since $\nu = 0$. However, the ZF equalization leads to noise enhancement unlike MMSE receivers.
\subsection{Variance of filter distortion due to multipath channel}
We can write the variance of the interference caused by filter distortion due to multipath channel from (\ref{5.15}) and (\ref{5.12}) as
\begin{eqnarray}\label{5.18}
\boldsymbol\gamma_{fd,m}\!\!\!\!&=&\!\!\!\!\mathcal {E}\|\boldsymbol{\psi}_{fd,m}\|^2 = \mathcal {E}\|\mathbf E\mathcal{F}\mathbf R_m\mathbf P_m^H \mathbf o_{fd}\|^2,\nonumber \\
\!\!\!\!&=&\!\!\!\! \mathcal {E}[\mathbf E\mathcal{F}\mathbf R_m\mathbf P_m^H \mathbf o_{fd}\mathbf o^H_{fd}\mathbf P_m\mathbf R_m^H\mathcal{F}^H\mathbf E^H],\nonumber \\
\!\!\!\!&=&\!\!\!\! \mathbf E\mathcal{F}\mathbf R_m\mathbf P_m^H \mathcal{E}[\mathbf o_{fd}\mathbf o^H_{fd}]\mathbf P_m\mathbf R_m^H\mathcal{F}^H\mathbf E^H,\nonumber \\
\!\!\!\!&=&\!\!\!\! \mathbf E\mathcal{F}\mathbf R_m\mathbf P_m^H \boldsymbol \alpha_{fd}\mathbf P_m\mathbf R_m^H\mathcal{F}^H\mathbf E^H,
\end{eqnarray}
Using (\ref{5.11}), we can determine $\boldsymbol\alpha_{fd}=\mathcal{E}[\mathbf o_{fd}\mathbf o^H_{fd}]$ as follows
\begin{eqnarray}\label{5.19}
\boldsymbol\alpha_{fd}\!\!\!\!&=&\!\!\!\! \mathcal {E}\Big[\Big\{ \sum_{l=0}^{L-1} \rho_l \mathbf Z_{l}\Delta\mathbf{P}^{\downarrow l}\mathbf{b}_e^{\downarrow l}\Big\}\{ \sum_{l=0}^{L-1} \rho_l \mathbf Z_{l}\Delta\mathbf{P}^{\downarrow l}\mathbf{b}_e^{\downarrow l}\Big\}^H\Big],\nonumber \\
\!\!\!\!&=&\!\!\!\! \sum_{l=0}^{L-1}\rho^2_l\mathcal {E}[\mathbf Z_{l}\Delta\mathbf{P}^{\downarrow l}\mathbf{b}_e^{\downarrow l}\mathbf{b}_e^{\downarrow lH}\Delta\mathbf{P}^{\downarrow lH}\mathbf Z^H_{l}]\nonumber \\
&=&\!\!\!\! \sum_{l=0}^{L-1}\rho^2_l\mathcal {E}[\mathbf Z_{l}\Delta\mathbf{P}^{\downarrow l}\mathbf X_l\mathcal F^H\mathbf D_m \mathbf s_m\mathbf s_m^H\mathbf D_m^H\mathcal F\mathbf X_l^H\Delta\mathbf{P}^{\downarrow lH}\mathbf Z^H_{l}]\nonumber \\
\end{eqnarray}
From (\ref{5.10}), we know that $\mathcal{E}\{\mathbf Z_l\mathbf Z_l^H\}\!\!=\!\!1$ since $z_l \in \mathbb{C}\mathcal{N}(0,1)$ also $\mathcal{E}\{\mathbf s_m\mathbf s_m^H\}=\delta^2\mathbf I$ and $\|\mathbf X_l\mathcal F^H\mathbf D_m\mathbf D_m^H\mathcal F\mathbf X_l^H\|_n=\frac{k}{n}\|\mathbf I_{N\times N}\|_n$, consequently
\begin{eqnarray}\label{5.20}
\boldsymbol\alpha_{fd}\!\!\!\!&=&\!\!\!\!\frac{k}{n}\delta^2\sum_{l=0}^{L-1}\rho^2_l\textrm{Tr}\{\Delta\mathbf{P}^{\downarrow l}\Delta\mathbf{P}^{\downarrow lH}\},\nonumber \\
\!\!\!\!&=&\!\!\!\!\frac{k}{n}\delta^2\sum_{l=0}^{L-1}\rho^2_l T^{\downarrow l},
\end{eqnarray}
where $T^{\downarrow l}\!\!=\!\! \textrm{Tr}[\Delta\mathbf{P}^{\downarrow l}\Delta\mathbf{P}^{\downarrow lH}]$. Since $T^{\downarrow l}$ is a scalar, $\boldsymbol\alpha_{fd}$ is also a scalar. Now substituting (\ref{5.20}) into (\ref{5.18}), yields
\begin{eqnarray}\label{5.21}
\boldsymbol \gamma_{fd,m}= \alpha_{fd}\mathbf E\mathcal{F}\mathbf R_m\mathbf P_m^H\mathbf P_m\mathbf R_m^H\mathcal{F}^H\mathbf E^H,
\end{eqnarray}
By taking the $n^{th}$ diagonal element of $\boldsymbol \gamma_{fd,m}$, we obtain
\begin{eqnarray}\label{5.22}
\gamma_{fd,m,n}\!\!\!\!&=&\!\!\!\!\alpha_{fd}\|\mathbf E\mathcal{F}\mathbf R_m\mathbf P_m^H\mathbf P_m\mathbf R_m^H\mathcal{F}^H\mathbf E^H\|_n,\nonumber \\
\!\!\!\!&=&\!\!\!\!\alpha_{fd}|E_n|^2\zeta_{m,n}.
\end{eqnarray}
where $\|\mathcal{F}\mathbf R_m\mathbf P_m^H\mathbf P_m\mathbf R_m^H\mathcal{F}^H\|_n=\zeta_{m,n}\|\mathbf I_{N\times N}\|_n$. \\
From (\ref{5.22}), we can see that the variance of filter distortion in FBMC/QAM-IM system (where $k<n$) is $\frac{k}{n}$ times the variance of filter distortion in FBMC/QAM system (where $k=n$). Hence, we can say that the inactive subcarriers do not contribute to the total interference in the FBMC/QAM-IM system since $\frac{k}{n} < 1$ in FBMC/QAM-IM system.
\subsection{Variance of IBI}
We can write the variance of the interference caused by IBI from (\ref{5.15}) and (\ref{5.12}) as
\begin{eqnarray}\label{5.23}
\boldsymbol\gamma_{IBI,m}\!\!\!\!&=&\!\!\!\!\mathcal {E}\|\boldsymbol{\psi}_{IBI,m}\|^2 
\!=\! \mathcal {E}\|\mathbf E\mathcal{F}\mathbf R_m\mathbf P_m^H\mathbf o_{IBI}\|^2,\nonumber \\
\!\!\!\!&=&\!\!\!\! \mathcal {E}[\mathbf E\mathcal{F}\mathbf R_m\mathbf P_m^H\mathbf o_{IBI}\mathbf o^H_{IBI}\mathbf P_m\mathbf R_m^H\mathcal{F}^H\mathbf E^H],\nonumber \\
\!\!\!\!&=&\!\!\!\! \mathbf E\mathcal{F}\mathbf R_m\mathbf P_m^H\mathcal {E}[\mathbf o_{IBI}\mathbf o^H_{IBI}]\mathbf P_m\mathbf R_m^H\mathcal{F}^H\mathbf E^H,\nonumber\\
\!\!\!\!&=&\!\!\!\! \mathbf E\mathcal{F}\mathbf R_m\mathbf P_m^H\boldsymbol\alpha_{IBI}\mathbf P_m\mathbf R_m^H\mathcal{F}^H\mathbf E^H,
\end{eqnarray}
where $\boldsymbol\alpha_{IBI} \!\!=\!\! \mathcal {E}[\mathbf o_{IBI}\mathbf o^H_{IBI}]$, now using (\ref{5.11}), we can determine $\boldsymbol\alpha_{IBI}$ as
\begin{eqnarray}\label{5.24}
\boldsymbol\alpha_{IBI}\!\!\!&=& \!\!\!\mathcal {E}\Big[\Big\{ \sum_{l=0}^{L-1} \rho_l \mathbf Z_{l}\mathbf{y}_{B,l}\Big\}\Big\{ \sum_{l=0}^{L-1} \rho_l \mathbf Z_{l}\mathbf{y}_{B,l}\Big\}^H\Big],\nonumber \\
\!\!\!&=&\!\!\! \mathcal {E}\Big[\sum_{l=0}^{L-1}\rho^2_l\mathbf Z_{l}\mathcal{E}\{\mathbf{y}_{B,l}\mathbf{y}^H_{B,l}\}\mathbf Z^H_{l}\Big],
\end{eqnarray}
Since $\mathbf Z_l$ has a complex Gaussian distribution i.e. $\mathbb{C}\mathcal{N}(0,1)$ and also $\mathbf Z_l$ and $\mathbf{y}_{B,l}$ are uncorrelated, we can write the above equation as follows
\begin{eqnarray}\label{5.25}
\boldsymbol\alpha_{IBI}=\sum_{l=0}^{L-1}\rho^2_l\mathcal{E}\{\mathbf{y}_{B,l}\mathbf{y}^H_{B,l}\},
\end{eqnarray}
$\mathcal {E}\{\mathbf{y}_{B,l} \mathbf{y}^H_{B,l}\}$ is dependent on the signal type of the last bock, where we assume it  is also occupied by an FBMC symbol with the same power, then we have 
\begin{eqnarray}\label{5.26}
\mathcal {E}\{\mathbf{y}_{B,l} \mathbf{y}^H_{B,l}\} \!\!\!\!&=&\!\!\!\! \mathcal{E}\|\mathbf {P}_{(l)}\mathbf {b}_{last}\|^2\!\!=\!\!\textrm{Tr}\big[\mathbf {P}_{(l)}\mathcal {E}\{\mathbf {b}_{last}{\mathbf {b}^H_{last}}\}\mathbf {P}^H_{(l)}\big],\nonumber \\
&=&\!\!\textrm{Tr}\big[\mathbf {P}_{(l)}\mathcal {E}\{\mathcal {F}^H\mathbf {D}_{last}\mathbf {s}_{last}\mathbf {s}_{last}^H\mathbf {D}_{last}^H\mathcal F\}\mathbf {P}^H_{(l)}\big],\nonumber \\
\!\!\!&=&\!\!\! \frac{k}{n}\delta^2\textrm{Tr}\big[\mathbf {P}_{(l)}\mathbf {P}^H_{(l)}\big]=\frac{k}{n}\delta^2\textrm{Tr}\big[\mathbf P^{corr}_{(l)}],\nonumber\\
\!\!\!&=&\!\!\!\frac{k}{n}\delta^2 P^{corr}_{(l)},
\end{eqnarray}
where $\mathbf{P}_{(l)} \!\!\!=\!\!\![\mathbf{P}_{(last-l)}; \mathbf 0_{(M+K-1)N-l \times MN}]$ in which $\mathbf{P}_{(last-l)}$ contains the last $l$-th rows of $\mathbf{P}$ also $\mathbf {b}_{last}$  is the symbol (after IDFT) in the last block and $\mathcal{E}\{\mathbf s_{last}\mathbf s_{last}^H\}=\delta^2\mathbf I$ and $\|\mathcal F^H\mathbf D_{last}\mathbf D_{last}^H\mathcal F\|_n=\frac{k}{n}\|\mathbf I_{N\times N}\|_n$. Substituting (\ref{5.26}) into (\ref{5.25}), we obtain
\begin{eqnarray}\label{5.27}
\alpha_{IBI} \!\!\!\!&=&\!\!\!\!\frac{k}{n}\delta^2\sum_{l=0}^{L-1}\rho^2_l P^{corr}_{(l)},
\end{eqnarray}
Since $P^{corr}_{(l)}$ is a scalar, $\alpha_{IBI}$ is also a scalar. Substituting (\ref{5.27}) into (\ref{5.23}), yields
\begin{eqnarray}\label{5.28}
\boldsymbol \gamma_{IBI,m}= \alpha_{IBI}\mathbf E\mathcal{F}\mathbf R_m\mathbf P_m^H\mathbf P_m\mathbf R_m^H\mathcal{F}^H\mathbf E^H,
\end{eqnarray}
By taking the $n^{th}$ diagonal element of $\boldsymbol \gamma_{IBI,m}$, we derive the MSE of IBI as
\begin{eqnarray}\label{5.29}
\gamma_{IBI,m,n}\!\!\!&=&\!\!\!\alpha_{IBI}\|\mathbf E\mathcal{F}\mathbf R_m\mathbf P_m^H\mathbf P_m\mathbf R_m^H\mathcal{F}^H\mathbf E^H\|_n,\nonumber\\
\!\!\!&=&\!\!\!\alpha_{IBI}|E_n|^2\zeta_{m,n}.
\end{eqnarray}
where $\|\mathcal{F}\mathbf R_m\mathbf P_m^H\mathbf P_m\mathbf R_m^H\mathcal{F}^H\|_n=\zeta_{m,n}\|\mathbf I_{N\times N}\|_n$. It should be noted that if we consider a sufficient guard interval between the data blocks then we can safely assume the inter-block interference to be negligible i.e., $\gamma_{IBI,m,n}=0$. Also from (\ref{5.29}), we can see that the variance of IBI in FBMC/QAM-IM system is also $\frac{k}{n}$ times the variance of IBI in FBMC/QAM system.
\subsection{Variance of Noise}
We can write the variance of the noise from (\ref{5.15}) and (\ref{5.12}) as
\begin{eqnarray}\label{5.30}
\boldsymbol\gamma_{noise,m}\!\!\!&=&\!\!\!\mathcal {E}\|\boldsymbol{\psi}_{noise,m}\|^2 = \mathcal {E}\|\mathbf E\mathcal{F}\mathbf R_m\mathbf P_m^H\mathbf n\|^2,\nonumber \\
\!\!\!&=&\!\!\! \mathcal {E}[\mathbf E\mathcal{F}\mathbf R_m\mathbf P_m^H\mathbf n\mathbf n^H\mathbf P_m\mathbf R_m^H\mathcal{F}^H\mathbf E^H],\nonumber \\
\!\!\!\!&=&\!\!\!\! \sigma^2\mathbf E\mathcal{F}\mathbf R_m\mathbf P_m^H\mathbf P_m\mathbf R_m^H\mathcal{F}^H\mathbf E^H
\end{eqnarray}
where $\mathcal{E}\{\mathbf n \mathbf n^H\}=\mathcal{E}\|\mathbf n\|^2=\sigma^2$ since $\mathbf{n}$ is Gaussian noise with each element having zero mean and variance $\sigma^2$. Taking the $n^{th}$ diagonal element of (\ref{5.30}), we have
\begin{eqnarray}\label{5.31}
\boldsymbol\gamma_{noise,m,n}\!\!\!\!&=&\!\!\!\!\sigma^2\|\mathbf E\mathcal{F}\mathbf R_m\mathbf P_m^H\mathbf P_m\mathbf R_m^H\mathcal{F}^H\mathbf E^H\|_n,\nonumber\\
\!\!\!\!&=&\!\!\!\!\sigma^2|E_n|^2\zeta_{m,n}.
\end{eqnarray}
where $\|\mathcal{F}\mathbf R_m\mathbf P_m^H\mathbf P_m\mathbf R_m^H\mathcal{F}^H\|_n=\zeta_{m,n}\|\mathbf I_{N\times N}\|_n$. Note that the $\zeta_{m,n}$ is the noise / interference enhancement factor which is introduced when we use an inverse filter matrix at the receiver. We can also see that the variance of noise in FBMC/QAM-IM system is the same as the variance of noise in FBMC/QAM system as it is independent of the active an inactive subcarrier selection. Hence, the performance improvement in FBMC/QAM-IM system comes from the reduction in the variance of interferences due to the use of index modulation.
\section{Improved Receiver for FBMC/QAM-IM}
The conventional receivers in FBMC/QAM system are used for detecting the $\mathcal M$-ary symbols to extract the transmitted information. However, the FBMC/QAM-IM receiver needs to detect the indexes of the active sub-carriers and also the corresponding information ($\mathcal M$-ary) symbol transmitted on those active subcarriers. To detect the active sub-carrier indexes and the $\mathcal M$-ary symbols on the active subcarriers, the received symbol vector $\mathbf{\hat s}_m$ is divided into $g$ groups by the detection group creator block i.e., $\mathbf {\hat s}_m = [\mathbf B_{m,1}, \mathbf B_{m,2}, ..., \mathbf B_{m,g}]^T\in\mathbb{C}^{N\times 1}$ as shown in Fig. \ref{fig-5.2}. Each sub-block $\mathbf B_{m,g} \in \mathbb{C}^{n\times 1}$ can now be detected by an optimum ML detector. The output of the detector is then used to extract the information embedded in the indexes of the active subcarriers as well as the constellation symbols transmitted on those active subcarriers. However, the optimal ML detector suffers from high complexity. In the following section, using the interference and noise power analysis presented in Section \ref{sec5.3}, we propose a low complexity detector based on the LLR approach. 
\subsection{Maximum likelihood (ML) Detector}
The ML detector is an optimum detector that considers all possible sub-block realizations by searching for all possible sub-carrier index combinations and signal constellation points to make a joint decision on the active indexes and the constellation symbols for each sub-block by minimizing the following metric
\begin{eqnarray}\label{5.32}
\mathbf {\hat A}_{m,\beta} = \arg \min\limits_{\mathbf A_{m,g} \in \Gamma}\big\|\mathbf B_{m,\beta}-\mathbf A_{m,\beta}\big\|^2
\end{eqnarray}
Thus, the ML detector chooses the sub-block $\mathbf A_{m,g} \in \Gamma$, where $\Gamma$ is the set of all the possible sub-blocks, that yields the smallest distance with the received sub-block $\mathbf B_{m,g}$ to estimate the transmitted sub-block. From the estimated sub-block $\mathbf {\hat A}_{m,\beta}$, the index bits and $\mathcal M$-ary symbol bits can then be decoded using the index decoder and $\mathcal M$-ary demodulator as shown in Fig. \ref{fig-5.2}. Although the ML detector can provide optimal performance but its complexity is unaffordable i.e., $\sim\mathcal O$($2^{p_1}\mathcal M^k$). Therefore, to reduce the complexity of the ML detector, we propose an improved LLR detector based on the interference and noise power analysis provided in Section \ref{sec5.3}. The complexity of a LLR detector is $\sim\mathcal O$($\mathcal M$) which makes it less complex than ML detector \cite{8287935}.
\subsection{Log-likelihood Ratio (LLR) Detector}
In this section we proposed an improved detector for FBMC/QAM-IM based on the interference analysis given in Section \ref{sec5.3}. A general LLR detector provides the logarithm of the ratio of a posteriori probabilities of the frequency domain symbols by considering the fact that their values can either be zero or non-zero depending upon the sub-carrier being active or inactive. To determine the status of any subcarrier being active or inactive, we can use the following ratio
\begin{eqnarray}\label{5.33}
\lambda_{m,n} = \log_e \frac{\sum_{\chi=1}^{\mathcal M}P(s_{m,n}=a_{\chi}\big|\hat s_{m,n})}{P(s_{m,n}=0\big|\hat s_{m,n})}
\end{eqnarray}
where $a_\chi\in \mathcal S$ and $\lambda_{m,n}$ is the LLR value of the $n^{th}$ subcarrier of $m^{th}$ symbol in a FBMC/QAM-IM block. It should be noted that a larger value of $\lambda_{m,n}$ means it is more probable that the $n^{th}$ subcarrier under consideration was selected by the index selection block at the transmitter or in other words the subcarrier was active. The LLR expression given in (\ref{5.33}) can be simplified by applying Bayes' formula as follows
\begin{eqnarray}\label{5.34}
\lambda_{m,n} \!\!\!\!\!&=&\!\!\!\!\! \log_e\! \frac{{\sum_{\chi=1}^{\mathcal M}\!P(\hat s_{m,n}\big|s_{m,n}\!=\!a_{\chi})P(s_{m,n}\!=\!a_{\chi})}/{P(\hat s_{m,n})}}{P(\hat s_{m,n}\big|s_{m,n}\!=\!0)P(s_{m,n}\!=\!0)/P(\hat s_{m,n})}\nonumber\\
\!\!\!\!\!&=&\!\!\!\!\!\log_e\! \frac{{\sum_{\chi=1}^{\mathcal M}P(\hat s_{m,n}\big|s_{m,n}=a_{\chi})P(s_{m,n}=a_{\chi})}}{P(\hat s_{m,n}\big|s_{m,n}=0)P(s_{m,n}=0)}
\end{eqnarray}
As we already know that 
\begin{eqnarray}\label{5.35}
\sum_{\chi=1}^{\mathcal M}P(s_{m,n}=a_{\chi}) = \frac{k}{n}
\end{eqnarray}
and,
\begin{eqnarray}\label{5.36}
P(s_{m,n}=0) = \frac{n-k}{n}
\end{eqnarray}
Using (\ref{5.35}) and (\ref{5.36}), we can update (\ref{5.34}) as
\begin{eqnarray}\label{5.37}
\lambda_{m,n} =\log_e \frac{k{\sum_{\chi=1}^{\mathcal M}P(\hat s_{m,n}\big|s_{m,n}=a_{\chi})}}{(n-k)P(\hat s_{m,n}\big|s_{m,n}=0)}
\end{eqnarray}
Eq. (\ref{5.37}) can be further simplified as
\begin{eqnarray}\label{5.38}
\lambda_{m,n} &=&\log_e(k) - \log_e(n-k) \nonumber \\
&&- \underbrace{\log_e\frac{\sum_{\chi=1}^{\mathcal M}P(\hat s_{m,n}\big|s_{m,n}=a_{\chi})}{P(\hat s_{m,n}\big|s_{m,n}=0)}}_{\theta_{m,n}}
\end{eqnarray}
According to (\ref{5.12}), the equalized symbol vector can be modeled as 
\begin{eqnarray}\label{5.39}
\mathbf {\hat s}_m=\mathbf s_m+\boldsymbol{\psi}_{tot,m}
\end{eqnarray}
where $\boldsymbol{\psi}_{tot,m}=\boldsymbol{\psi}_{resd,m}+\boldsymbol{\psi}_{fd,m}+\boldsymbol{\psi}_{IBI,m}+\boldsymbol{\psi}_{noise,m}$ is the sum of interference terms and the processed noise in the FBMC/QAM-IM system. It should be noted that the noise term $\boldsymbol{\psi}_{noise,m}$ is independent of all other terms and interference; the IBI contribution $\boldsymbol{\psi}_{IBI,m}$ is also independent of all other terms since the interference comes from the previous FBMC/QAM-IM block. However, the MMSE estimation bias error $\boldsymbol{\psi}_{resd,m}$ and filter distortion due to multipath channel $\boldsymbol{\psi}_{fd,m}$ are correlated since they both depend on the desired signal $\mathbf s_m$. A ZF equalizer can be used to avoid the MMSE estimation bias error and to have all the remaining interference terms and noise independent with each other. We can now write the third term i.e., $\theta_{m,n}$ in (\ref{5.38}) as follows
\begin{eqnarray}\label{5.40}
\theta_{m,n}&=&\!\!\!\log_e\frac{\sum_{\chi=1}^{\mathcal M}P(\hat s_{m,n}\big|s_{m,n}=a_{\chi})}{P(\hat s_{m,n}\big|s_{m,n}=0)}\nonumber \\
&=&\!\!\!\log_e\frac{\sum_{\chi=1}^{\mathcal M} \frac{1}{\pi\gamma_{tot,m,n}} \textrm{exp}\bigg({\frac{-|\hat s_{m,n}-a_\chi|^2}{\gamma_{tot,m,n}}}\bigg)}{\frac{1}{\pi\gamma_{tot,m,n}} \textrm{exp}\bigg({\frac{-|\hat s_{m,n}|^2}{\gamma_{tot,m,n}}}\bigg)}\nonumber \\
&=&\!\!\!\log_e\frac{\sum_{\chi=1}^{\mathcal M} \textrm{exp}\bigg({\frac{-|\hat s_{m,n}-a_\chi|^2}{\gamma_{tot,m,n}}}\bigg)}{\textrm{exp}\bigg({\frac{-|\hat s_{m,n}|^2}{\gamma_{tot,m,n}}}\bigg)}\nonumber \\
&=&\!\!\!\frac{-|\hat s_{m,n}|^2}{\gamma_{tot,m,n}}\!\!+\!\!\log_e\!\Bigg\{\!\!\sum_{\chi=1}^{\mathcal M} \textrm{exp}\bigg(\!\!{\frac{-|\hat s_{m,n}-a_\chi|^2}{\gamma_{tot,m,n}}}\!\!\bigg)\!\!\Bigg\}
\end{eqnarray}
Substituting (\ref{5.40}) into (\ref{5.38}) yields,
\begin{eqnarray}\label{5.41}
\lambda_{m,n} &=&\log_e(k) - \log_e(n-k) + \frac{|\hat s_{m,n}|^2}{\gamma_{tot,m,n}}\nonumber \\
&&+\log_e\Bigg\{\!\sum_{\chi=1}^{\mathcal M} \textrm{exp}\bigg({\frac{-|\hat s_{m,n}-a_\chi|^2}{\gamma_{tot,m,n}}}\bigg)\!\Bigg\}
\end{eqnarray}
where $\gamma_{tot,m,n}$ is the total noise plus interference power of the $n^{th}$ subcarrier of the $m^{th}$ symbol in a FBMC/QAM-IM block and can be calculated using (\ref{5.15}). After calculating the $n$ LLR values of each sub-block of the $m^{th}$ FBMC/QAM-IM symbol, the $k$ subcarriers in each sub-group which have maximum LLR value are assumed to be active. After detection of active sub-carrier indexes in each sub-group, the information is passed to the \textit{index decoder} which provides the estimate of the index selecting $p_1$ bits based on the indexes of the active sub-carriers in each sub-group. The $\mathcal M$-ary constellation symbols transmitted on each active sub-carriers is demodulated by the \textit{$\mathcal M$-ary demodulator} in a conventional manner to estimate the remaining $p_2$ bits. The \textit{bit combiner} block then combines $p_1$ and $p_2$ bits from all the sub-blocks to generate the transmitted bit vector $\hat T$ as shown in Fig. \ref{fig-5.2}.
\section{Simulation Results}
In this section we present the simulation results for MSE, output SINR and SIR in FBMC/QAM system with and without index modulation along with the BER performance comparison of index modulation based FBMC/QAM system and conventional FBMC/QAM system.
\begin{figure}[!htbp]
	\centering
	\begin{subfigure}[b]{0.5\textwidth}
		\includegraphics[width=\textwidth]{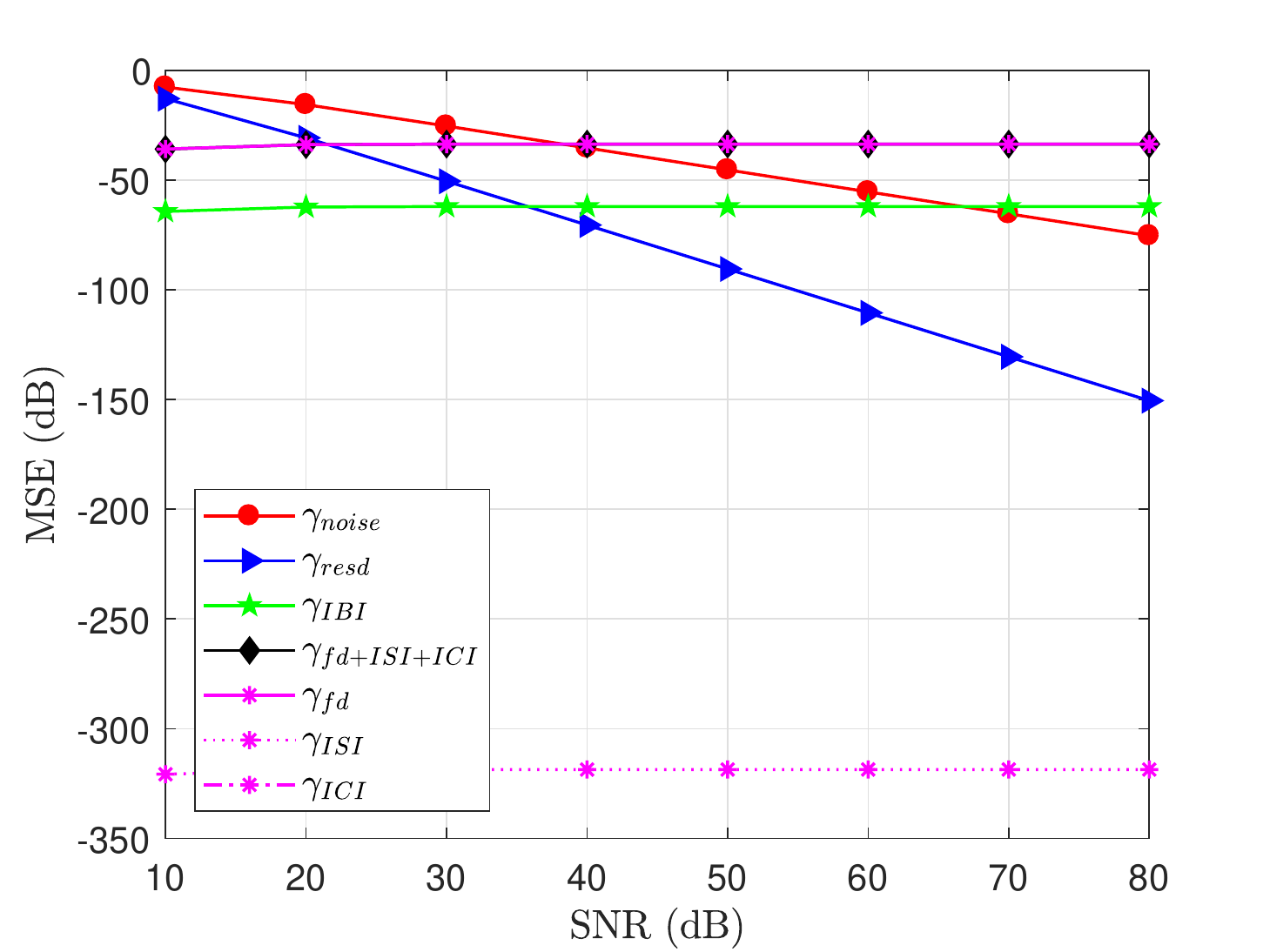}
		\caption{Individual MSE Components (FBMC/QAM)}
		\label{fig-5.3a}
	\end{subfigure}
	~ 
	\begin{subfigure}[b]{0.5\textwidth}
		\includegraphics[width=\textwidth]{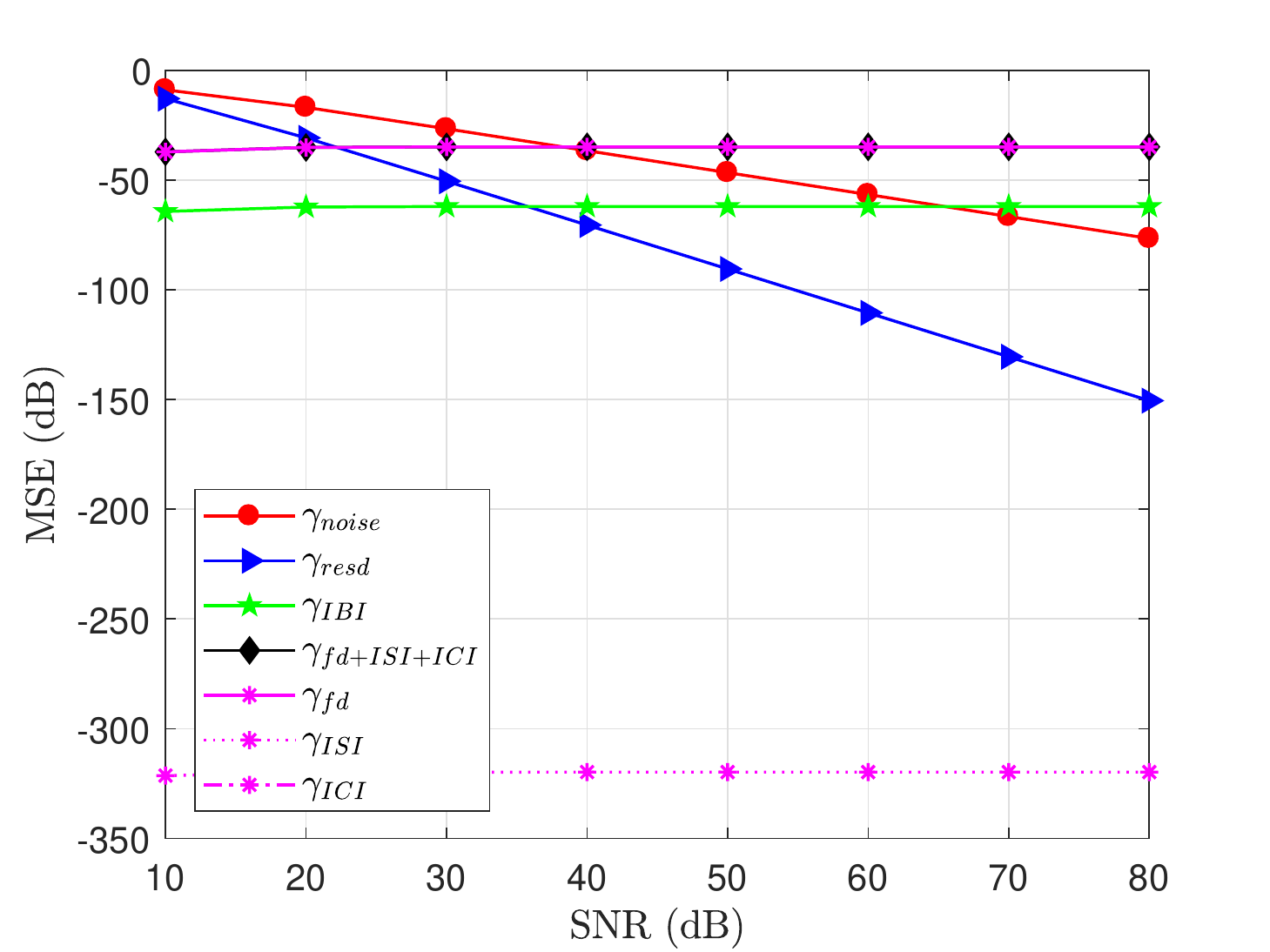}
		\caption{Individual MSE Components (FBMC/QAM-IM)}
		\label{fig-5.3b}
	\end{subfigure}
	\caption{Interference components in FBMC/QAM and FBMC/QAM-IM (4,3,QPSK)}\label{fig-5.3}
\end{figure}
\subsection{MSE and output SINR}
\begin{figure}[htbp]
	\centering
	\begin{subfigure}[b]{0.5\textwidth}
		\includegraphics[width=\textwidth]{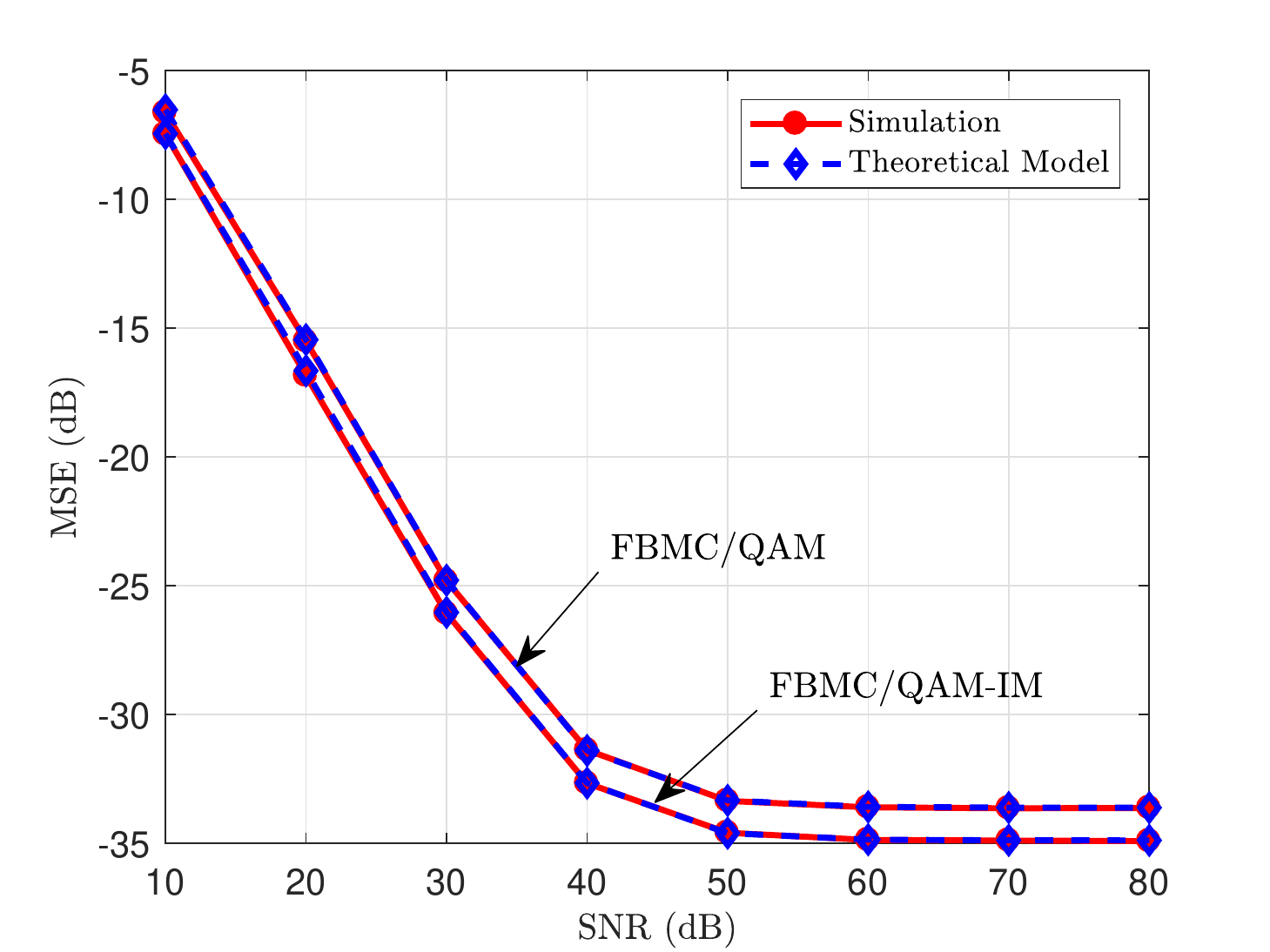}
		\caption{MSE performance comparison}
		\label{fig-5.4a}
	\end{subfigure}
	~ 
	\begin{subfigure}[b]{0.5\textwidth}
		\includegraphics[width=\textwidth]{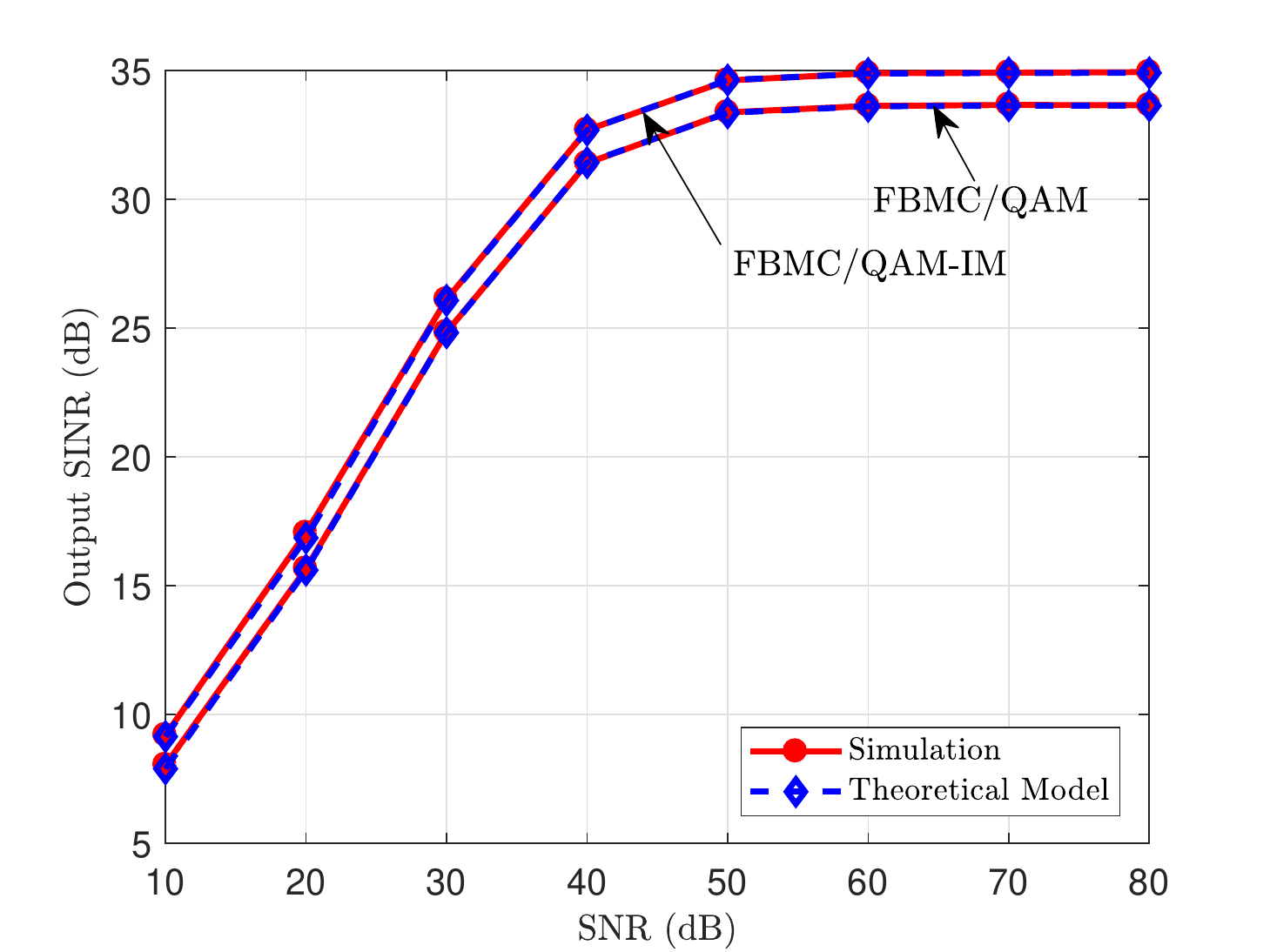}
		\caption{Output SINR performance comparison}
		\label{fig-5.4b}
	\end{subfigure}
	\caption{Performance comparison of FBMC/QAM and FBMC/QAM-IM (4,3,QPSK)}\label{fig-5.4}
\end{figure}

Since we know that not all of the subcarriers in FBMC/QAM-IM system are active i.e., $k<n$ unlike conventional FBMC/QAM system where $k=n$. In this case the interference power will be smaller than the conventional FBMC/QAM system. For our analysis, we consider $n=4$ and $k=3$ FBMC/QAM-IM system with QPSK as the $\mathcal M$-ary signal constellation. The individual interference terms like noise, residue from the MMSE equalization, IBI and filter distortion due to multipath channel in the proposed FBMC/QAM and FBMC/QAM-IM systems are derived in Section \ref{sec5.3} and the results are presented in Fig. {\ref{fig-5.3a}} and Fig. {\ref{fig-5.3b}} respectively. The results show the power of each interference component that is affecting the multicarrier system. It can be seen that the contribution of ICI and ISI (intrinsic interference) is quite insignificant with the use of inverse filter at the receiver i.e., ISI is around -320dB and ICI cannot be even displayed on the same scale. However, the system is still affected by residue from the MMSE equalization, IBI and filter distortion due to multipath channel. Since some of the subcarriers in IM based FBMC/QAM system are in-active, they will not contribute to these residual interferences. As a result, the interference level would be smaller compared to conventional FBMC/QAM system. The performance in terms of total MSE and output SINR in a FBMC/QAM system with and without index modulation is presented in Fig. \ref{fig-5.4}. It can be seen from Fig. \ref{fig-5.4a} that the MSE in FBMC/QAM system is improved with the use of IM. The improvement gain depends on the selection of $n$ and $k$ values. In this case we have considered a $n=4$ and $k=3$ which result in a gain around $\mathrm{10\log_{10}}({\frac{n}{k}})$ i.e., $\sim$ 1.25dB. The improvement in MSE performance can be enhanced by a higher $\frac{n}{k}$ ratio. 
\\The output SINR of the system also improves with the use of IM as can be seen from the Fig. \ref{fig-5.4b}. It can also be confirmed that the interference terms in the system model give in (\ref{5.12}) completely matches with the simulation results, which verifies the accuracy of the derived analytical model.
\subsection{SIR Performance}
\begin{figure}[h]
	\begin{center}
		\includegraphics[scale=0.65]{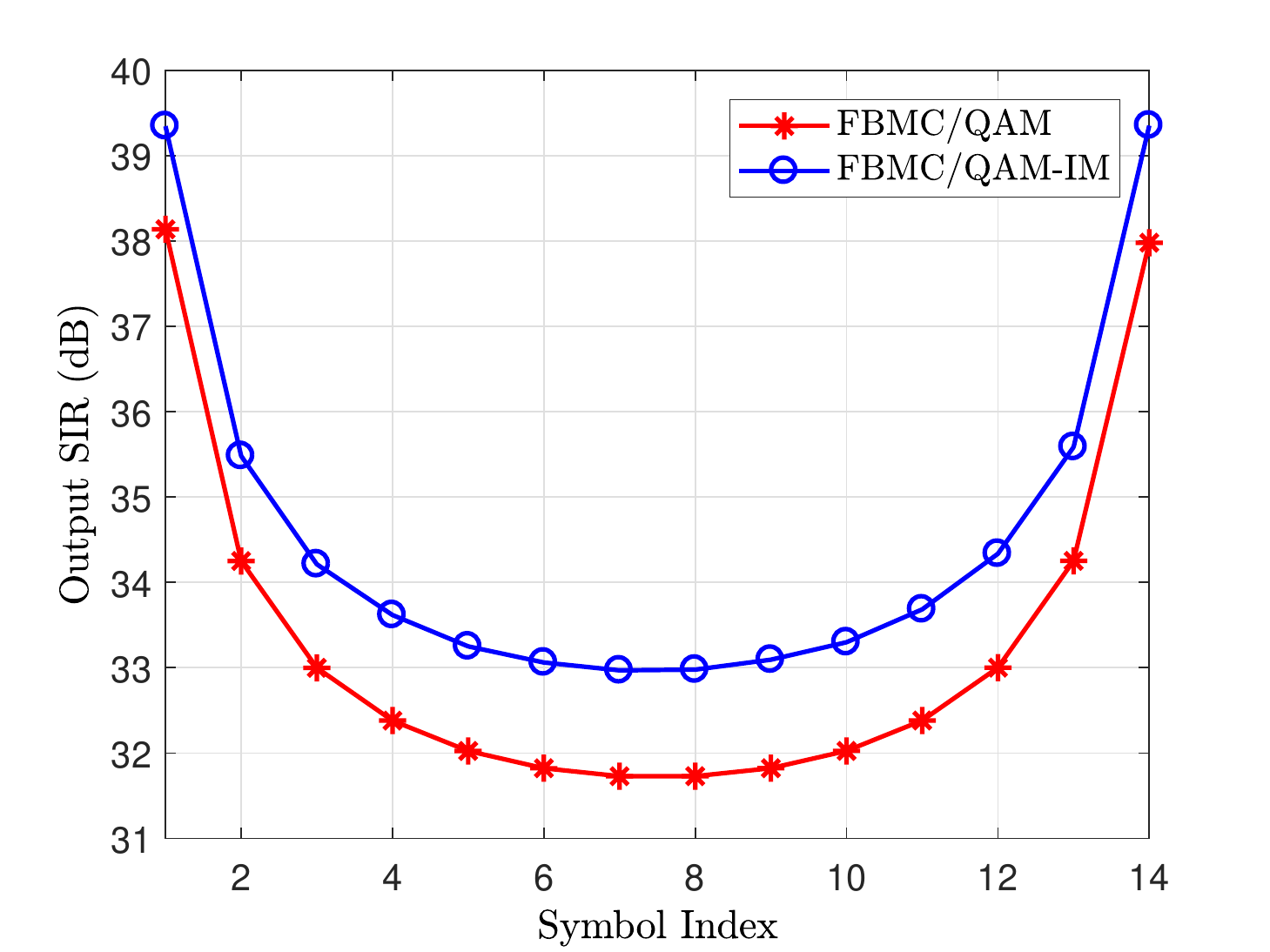}
		\caption{Output SIR performance comparison of FBMC/QAM and FBMC/QAM-IM (4,3,QPSK) with inverse filter}\label{fig-5.5}
	\end{center}
\end{figure}

The output SIR of FBMC/QAM system with and without the IM is presented in Fig. \ref{fig-5.5}. As we have discussed earlier that since some of the subcarriers are inactive in FBMC/QAM-IM subblock. Their power can either be saved to improve the energy efficiency of the system or it can be reallocated to the active subcarriers in a subgroup to improve the system BER performance. In our case we have considered the later option and distribute the power of inactive subcarriers to the active subcarriers. 
We have already established in Section \ref{sec5.3} that the interference power in FBMC/QAM-IM system has been reduced by $\mathrm{10\log_{10}}({\frac{n}{k}})\sim 1.25dB$ compared to conventional FBMC/QAM system. We can also see that the interference power is affecting the middle symbols more than the symbols at the edges of the FBMC/QAM block. The main reasons for this behavior is the intrinsic interference in FBMC/QAM system. As we know that symbols in FBMC/QAM system overlap each other both in time and frequency domain due to per subcarrier filtering. So it is obvious that the symbols at the edges of the block will experience less interference from the neighboring symbols compared to the symbols in the middle. Secondly, it has been already established that the use of inverse filter enhances the residue interferences in the FBMC/QAM system and that the enhancement factor affects the middle symbols more than the symbols at the edges as discussed in \cite{2017arXiv171108850Z}. It can be seen in Fig. \ref{fig-5.5} that the use of IM with FBMC/QAM improves the SIR of the system by reducing the variance of the interferences existing in the conventional FBMC/QAM system.
\subsection{BER Performance}\label{sec5.4.2}
\begin{figure}[h]
	\begin{center}
		\includegraphics[scale=0.65]{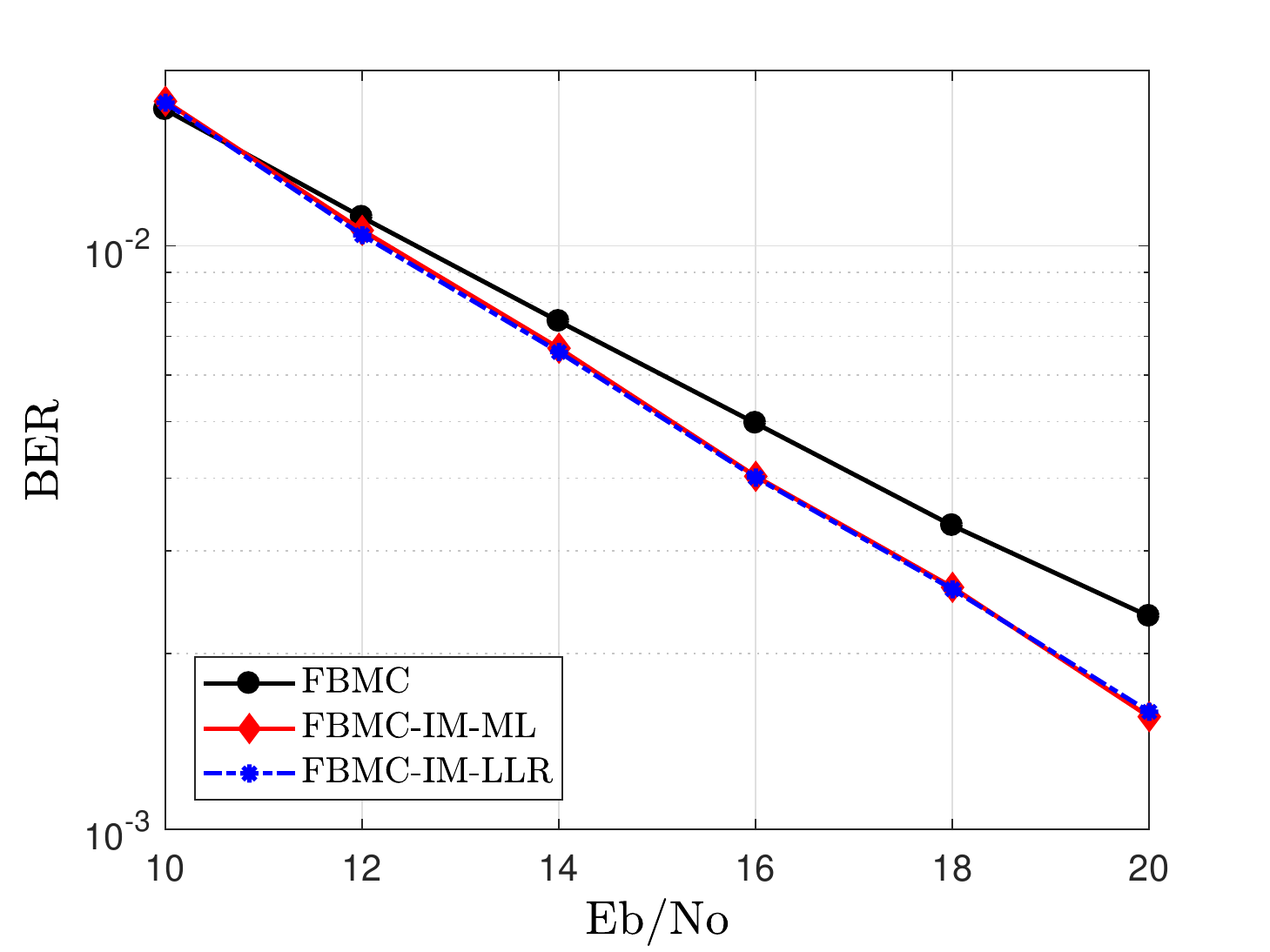}
		\caption{BER performance of FBMC/QAM system with and without IM}\label{fig9}
	\end{center}
\end{figure}

The results for the BER performance of FBMC/QAM system with and without IM are presented in Fig. {\ref{fig9}}. 
For FBMC/QAM-IM system, we have selected $(n,k)$ as $(4,3)$, which means that the SE of FBMC/QAM-IM is the same as the conventional FBMC/QAM system. It can be seen from the results that FBMC/QAM with IM has better performance compared to conventional FBMC/QAM system due to the presence of relatively lower interference. Since, FBMC/QAM-IM reduces the effect of residual interference at the receiver and also the power from inactive subcarriers are reallocated to the active carriers, the system can provide improved BER performance compared to its conventional counterpart. It can also be seen that the proposed LLR detector exhibit same performance as ML detector but with much lower complexity. \\
In the light of all the results, the improved performance of FBMC/QAM with IM compared to conventional FBMC/QAM systems makes it a suitable candidate for next generation wireless applications.
\section{Conclusion}
We have evaluated the performance of IM based QAM-FBMC system to highlight the potential of combining an emerging 5G modulation technique with our proposed FBMC/QAM system in \cite{2017arXiv171108850Z}. We first derived a mathematical model of the IM based QAM-FBMC system along with the derivation of interference terms at the receiver due to channel distortions and the intrinsic behavior of the transceiver model. We have shown that the interference power in FBMC/QAM-IM is smaller compared to that of conventional FBMC/QAM system as some subcarriers are inactive in IM based FBMC/QAM system. We then evaluated the performance of FBMC/QAM-IM in term of MSE and SINR and the results are compared with that of the conventional FBMC/QAM system. The results show that combining IM with FBMC/QAM can improve the system performance since the inactive subcarriers do not contribute to the overall interference in the system. The SIR performance of a FBMC/QAM block with and without IM is also presented. The results show the effect of interference on each FBMC/QAM symbol in a block. It can seen that the interference is higher for symbols in the middle of the block. At the end, BER performance of FBMC/QAM system with and without IM is presented and it can be seen that since the power from inactive subcarriers are reallocated to the active subcarriers, the FBMC/QAM-IM has shown improved performance compared to conventional FBMC/QAM system.
\bibliographystyle{IEEEtran}
\bibliography{IEEEabrv,varUD}
\end{document}